\documentclass[aps,pra,twocolumn,superscriptaddress]{revtex4-1}

\usepackage{graphicx}
\usepackage{hyperref}
\usepackage{amsmath}
\usepackage{amssymb}
\usepackage{epstopdf}
\usepackage{multirow}
\usepackage{threeparttable}
\usepackage[usenames]{color}

\begin{document}

\title{Mid-infrared single-photon computational temporal ghost imaging}
\author{Wen Zhang}
\affiliation{State Key Laboratory of Precision Spectroscopy, and Hainan Institute, East China Normal University, Shanghai 200062, China}

\author{Kun Huang}
\email{khuang@lps.ecnu.edu.cn}
\affiliation{State Key Laboratory of Precision Spectroscopy, and Hainan Institute, East China Normal University, Shanghai 200062, China}
\affiliation{Chongqing Key Laboratory of Precision Optics, Chongqing Institute of East China Normal University, Chongqing 401121, China}
\affiliation{Collaborative Innovation Center of Extreme Optics, Shanxi University, Taiyuan, Shanxi 030006, China}

\author{Xu Wang}
\affiliation{State Key Laboratory of Precision Spectroscopy, and Hainan Institute, East China Normal University, Shanghai 200062, China}

\author{Ben Sun}
\affiliation{State Key Laboratory of Precision Spectroscopy, and Hainan Institute, East China Normal University, Shanghai 200062, China}

\author{Jianan Fang}
\affiliation{State Key Laboratory of Precision Spectroscopy, and Hainan Institute, East China Normal University, Shanghai 200062, China}

\author{Yijing Li}
\affiliation{State Key Laboratory of Precision Spectroscopy, and Hainan Institute, East China Normal University, Shanghai 200062, China}

\author{Heping Zeng}
\email{hpzeng@phy.ecnu.edu.cn}
\affiliation{State Key Laboratory of Precision Spectroscopy, and Hainan Institute, East China Normal University, Shanghai 200062, China}
\affiliation{Chongqing Key Laboratory of Precision Optics, Chongqing Institute of East China Normal University, Chongqing 401121, China}
\affiliation{Shanghai Research Center for Quantum Sciences, Shanghai 201315, China}
\affiliation{Chongqing Institute for Brain and Intelligence, Guangyang Bay Laboratory, Chongqing, 400064, China}

\begin{abstract}
The capture of transient optical waveforms is critical to reveal dynamical phenomena in various fields. However, fast and sensitive mid-infrared (MIR) measurements are typically limited by processing bandwidth and detection sensitivity of conventional infrared detectors. Here, we propose and implement a computational temporal ghost imaging system, which favors high-speed and high-sensitivity characterization of MIR temporal objects. The core process relies on high-fidelity nonlinear optical transduction for facilitating both the programmable structured illumination and frequency upconversion detection based on the high-performance near-infrared light modulator and detector, respectively. Consequently, the correlation between the recorded integral upconversion intensity and the designated encoding patterns allows one to reconstruct the MIR profiles with a temporal resolution of 80 ps, well beyond the intrinsic bandwidth or timing jitter of the involved detectors. Moreover, a record-high detection sensitivity is manifested by recovering single-photon MIR waveforms with an incident flux below 0.1 photon/bit. Additionally, faithful reconstructions at sub-Nyquist sampling rates are demonstrated using the compressive sensing algorithm, which can reduce the data acquisition time by over 90\%. The presented paradigm features high timing precision, single-photon sensitivity, and efficient data sampling, which could be extended into far-infrared or terahertz regions to address pressing demands in fast and sensitive sensing.
\end{abstract}

\maketitle

\section{Introduction}
High-speed detection of optical signals is essential across a wide range of applications that require the rapid and accurate processing of light-based information, such as fiber-optic communication \cite{Assefa2010Nature}, photonic computing \cite{Shastri2021NP}, ultrafast spectroscopy \cite{Mahjoubfar2017NP}, and time-of-flight imaging \cite{Kim2021NN}. In these applications, a high temporal resolution of the light detection is usually required to realize low time latency, fast data rate, and enhanced depth resolution. Nowadays, the detection bandwidth for optical detectors at telecom wavelengths has reached impressive values over 100 GHz \cite{Shi2024NP, Wey1995JLT}, with an aim to fulfill the expanding demand for data communication capacity in big data, cloud computing and artificial intelligence \cite{Shastri2021NP}. However, the detection performances in other spectral regions are relatively lagging behind, especially in the mid-infrared (MIR) range \cite{Razeghi2014RPP}. Specifically, the speed response of conventional MIR detectors based on mercury-cadmium-telluride (MCT) alloys is severely limited by the long carrier lifetime \cite{Razeghi2014RPP}, which typically results in a bandwidth below GHz for commercially available products. Recently, MIR uni-traveling carrier (UTC) photodetectors based on InAs/InAsSb type-II superlattice \cite{Huang2022EDL} and quantum-well infrared photodetectors (QWIPs) based on group III-V materials \cite{Palaferri2018Nature} have been implemented to offer an operation bandwidth beyond 10 GHz, albeit with a smaller detection responsivity due to the reduced thickness of the constituent absorption layer. Additionally, narrow-bandgap MIR detectors have long been plagued by dark current and thermal perturbation, which severely limits the detection sensitivity at the room-temperature condition \cite{Wang2019Small}. Consequently, enhanced sensitivity usually resorts to the cryogenic operation, like superconducting nanowire detectors \cite{Taylor2023Optica, Pan2022OE, Chen2021SB}. Notably, emerging low-dimensional materials and novel nanophotonic structures have been investigated to realize room-temperature MIR detectors \cite{Wu2021NR, Xue2024LSA, Mu2024NP}, yet the attainable sensitivity is currently far away from the single-photon level. Therefore, it remains a long-sought-after goal for direct MIR detectors to achieve high sensitivity and high speed, as well as being compatible with room-temperature operation.

Alternatively, the temporal ghost imaging (TGI) technique provides an appealing solution to mitigate the stringent bandwidth requirement for high-speed signal detection \cite{Chen2013OE, Ryczkowski2016NP}. In this scenario, a reference beam without `seeing' the temporal object is dynamically modulated with random or structured patterns, while a test beam after passing through the target is detected by a non-temporally-resolved `bucket' detector \cite{Ryczkowski2016NP}. Consequently, the fast temporal object can be reconstructed with a slow integrating photodiode based on the intensity correlation between the two beams, although neither of them independently carries the temporal characteristics of the object \cite{Tian2020PRAp, Wu2023AO}. Fundamentally, the TGI is analogous to the spatial counterpart from the perspective of the space-time duality \cite{Patera2018PRA}, which favors a strong immunity to the waveform distortion due to noises and perturbations along the path between the object and detector \cite{Erkmen2010AOP, Moreau2018LPR}. To date, TGI has been demonstrated by using a variety of light sources, including spontaneous parametric downconversion (SPDC) photon pairs \cite{Denis2017JO, Dong2016SR}, chaotic laser sources \cite{Ryczkowski2016NP, Wu2020OE}, and thermal light sources \cite{Devaux2017JO, Yao2018OL}. Meanwhile, enhanced TGI variants based on temporal magnification \cite{Ryczkowski2017APLP} and differential measurement \cite{Oka2017APL} have been investigated to promote the temporal resolution and signal-to-noise ratio. In particular, computational TGI uses a pre-programmed light intensity modulator to generate a sequence of known illumination probe patterns, which removes the need for detecting the reference beam \cite{Devaux2016Optica, Xu2018OE, Chen2021OL}. These developments in TGI have benefited a number of applications, such as long-distance communication \cite{Chen2021OL, Yao2018OL}, information cryptography \cite{Jiang2017SR}, quantum device evaluation \cite{Wu2019OL}, and high-speed spectroscopy \cite{Janassek2018PRAp}. However, the aforementioned instantiations of TGI have mostly been restricted in the visible or near-infrared (NIR) regime due to the relatively easier accessibility of high-speed detectors or modulators. To leverage the full potential of TGI, it is desirable to extend the operation wavelength into the MIR region, as being fueled by the demand for fast infrared sensing in various fields such as high-throughput free-space communication \cite{Zou2022NC}, time-resolved photoluminescence spectroscopy \cite{Julsgaard2020PR}, and high-resolution depth tomography \cite{Fang2023LSA}.

In this context, the nonlinear frequency conversion strategy has been exploited in the TGI architecture, where the technical challenge in MIR detection can be mitigated by translating the infrared radiation into visible or NIR  spectral region \cite{Wu2019Optica}. In this approach, a two-color scheme is implemented by converting the probe temporal intensity pattern into the second-harmonic replica via a nonlinear crystal \cite{Wu2019Optica}. The involved wavelength conversion allows the shift of the probe laser into a spectral region where ultrafast detectors are available, which facilitates the imaging of a temporal object at 2 $\mu$m. However, this technique requires an infrared chaotic source with random intensity fluctuations, and relies on a large number of measurement realizations to obtain a high-contrast profile of the temporal object. Additionally, the conversion efficiency of the harmonic generation depends on the laser intensity, which would impose difficulty in preparing high-power sources in the MIR spectral region. To this end, MIR computation TGI has been proposed and implemented based on the pre-programmed structured illumination \cite{Wu2024LSA}, where the required MIR probe patterns are generated by the difference frequency generation (DFG) process from the fast-modulated NIR light. The presented method could in principle generate probe patterns at arbitrary wavelengths, and significantly improve the reconstruction speed by using the invertible Hadamard matrix \cite{Edgar2019NP}. However, in these reported TGI modalities, the MIR measurement for the test beam through the temporal object has still been conducted by using conventional narrow-bandgap infrared detectors, thus precluding the realization of ultra-sensitive operation at the single-photon level. So far, MIR single-photon computational TGI has not yet been demonstrated, which urgently calls for effective solutions to realize high-speed intensity modulation and high-sensitivity photon counting in the MIR region \cite{Kutas2022AQT}.

In this work, we devise and implement an ultra-sensitive MIR computational TGI system, which uniquely integrates both the nonlinear downconversion and upconversion units for performing high-fidelity pattern generation and single-photon photon detection at MIR wavelengths. In the illumination part, the fast modulated NIR light at 1.55 $\mu$m is nonlinearly converted into a sequence of MIR probe intensity patterns at 3.4$\mu$m, which relies on the DFG process within a nonlinear crystal under a continuous-wave pumping at 1.064 $\mu$m. Then, the generated MIR probe passes through a temporal object before being directed into a frequency upconversion detection unit. The upconversion detector favors a high efficiency and a low noise based on the cavity-enhanced sum-frequency generation (SFG) process, which provides the missing instrumentation for sensitive MIR detection in previous TGI systems. Finally, the upconverted waveforms at 0.81 $\mu$m can be recorded by high-performance room-temperature silicon detectors. The measured integral intensities in combination with the designated modulated patterns allow one to retrieve the binary-type or gray-scale temporal objects. The achieved temporal resolution is determined by the 12.5-GHz bandwidth of the used NIR modulator in the experiment, which can be further enhanced by resorting to more advanced commercially available modulators over 100 GHz. 

Moreover, the presented upconversion TGI system equipped with a single-photon counting module (SPCM) exhibits a record-high detection sensitivity, which permits the recovery of extremely weak MIR waveforms with a light flux below 0.1 photon/bit for the incident encoding sequence. Thanks to the high timing precision of the prepared probes, the achieved temporal resolution of 80 ps is no longer limited by the intrinsic timing jitter of the single-photon detector, which contrasts with conventional time-correlated detection scheme. Notably, a compressive sensing algorithm is used to facilitate the sub-Nyquist sampling strategy, which significantly reduces the data acquisition time by 90\% while maintaining a high recovery quality with a Pearson correlation coefficient over 85\%. Therefore, the presented MIR TGI system features picosecond timing resolution, single-photon sensitivity, and efficient data sampling, which would open up novel possibilities in fast and sensitive analysis for infrared temporal waveforms in low-light-level scenarios, for instance long-haul optical information transfer, photon-starved fluorescence lifetime measurement, and non-invasive time-domain infrared spectroscopy.\\
\newline

\begin{figure*}[t!]
\includegraphics[width=0.95\textwidth]{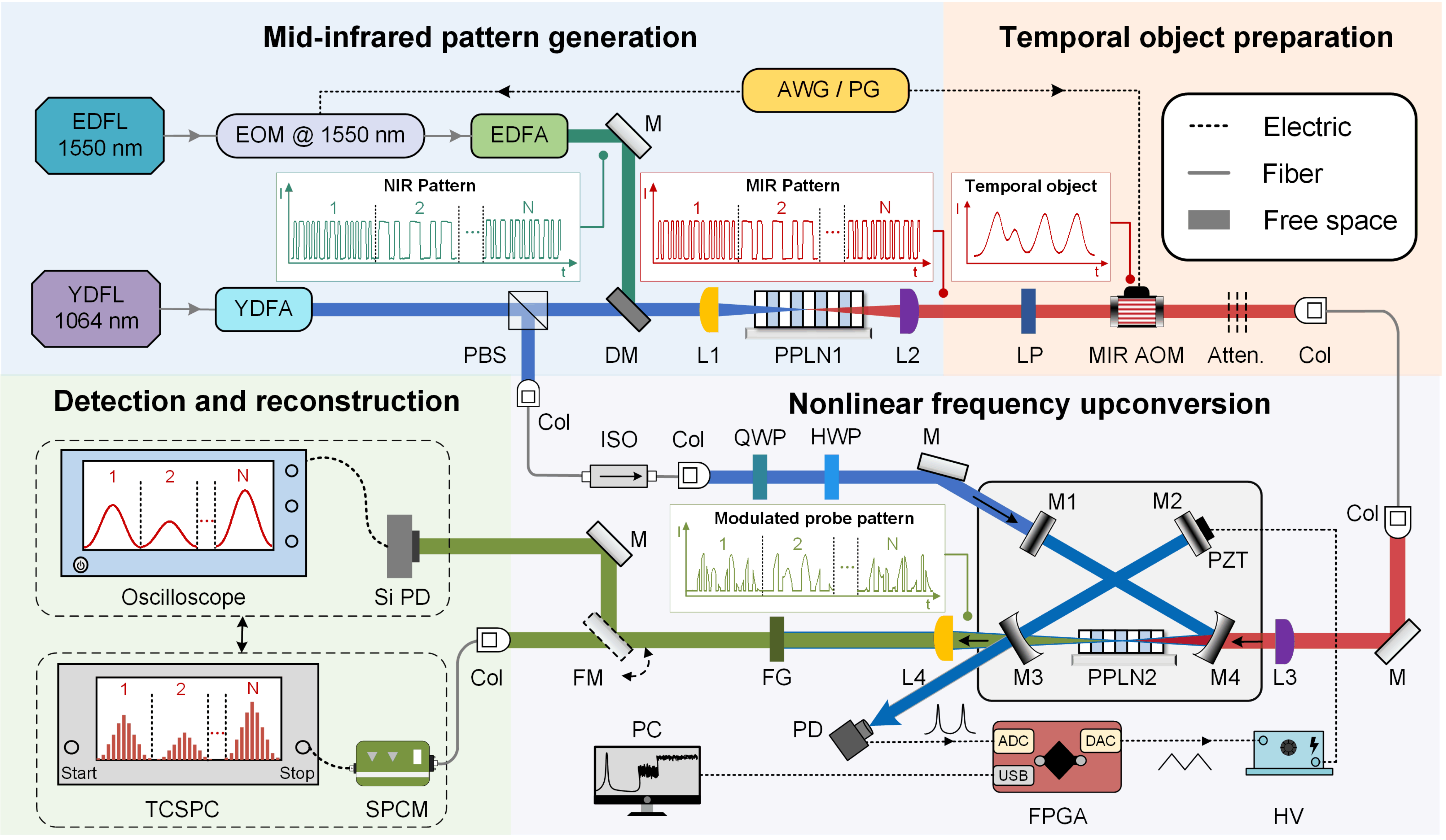}
\caption{Experimental setup for the MIR computational TGI system. An Er-doped fiber laser (EDFL) at 1550 nm and an Yb-doped fiber laser (YDFL) at 1064 nm are used to prepare the MIR patterns through the difference-frequency generation (DFG) process in a periodically poled lithium niobate (PPLN1) crystal. The continuous-wave  EDFL is temporally modulated by a fast electro-optic modulator (EOM) that is driven by a sequence of pre-programmed electrical patterns from an arbitrary waveform generator (AWG) or a high-speed pulse generator (PG). The MIR probe at 3.4 $\mu$m then passes through a temporal object emulated by an acousto-optic modulator (AOM). The intensity transmission profile of the temporal object can be controlled by the AWG or PG. Finally, the transmitted MIR light is sent into a pump-enhanced upconverter based on an external optical cavity. The involved sum-frequency generation (SFG) is performed by another nonlinear crystal (PPLN2) placed within the cavity. The cavity is actively stabilized by adapting the position of a mirror mounted on the piezoelectric actuator (PZT). The PZT is driven by a high-voltage (HV) source with a feedback seed from a digital locking unit based on the field-programmable gate array (FPGA). After a group of spectral filters, the upconverted signal at 810 nm is detected by a silicon-based photodiode (PD) or a single-photon counting module (SPCM). The outputs from the two types of detectors are recorded by a digital oscilloscope or a time-correlated single-photon counting (TCSPC) device. The correlation between the recorded integral signals and the predefined structured patterns allows the recovery of the MIR temporal object. M: silver mirror; DM: dichroic mirror; ISO: isolator; HWP: half-wave plate; QWP: quarter-wave plate; Atten: neutral density attenuator; Col: fiber collimator; L: Lens; LP: long-pass filter; FG: filtering group; FM: flipping mirror; PC: computer.}
\label{fig1}
\end{figure*}

\section{Experimental setup}
Figure \ref{fig1} illustrates the experimental setup for the MIR computational TGI system. The whole system relies on the frequency nonlinear conversion to implement the fast probe preparation and sensitive light detection for the MIR radiation. In the experiment, the involved light sources originate from an Er-doped fiber laser (EDFL) at 1550 nm and an Yb-doped fiber laser (YDFL) at 1064 nm. The EDFL delivers continuous-wave (CW) linearly polarized light with an average power of 27 mW. The CW beam is fast modulated by an electro-optic modulator (EOM) with an operation bandwidth of 20 GHz. The EOM is driven by a programmed sequence of electrical signals from an arbitrary waveform generator (AWG, RIGOL, DG4162) or a pulse pattern generator (Anritsu, MP1763C). The latter signal generator with a bandwidth of 12.5 GHz can facilitate a high-speed binary encoding with a temporal resolution of 80 ps. Then, the NIR modulated light is spatially combined with the pump beam from the YDFL via a dichroic mirror. The mixed beams are focused by a lens into a periodically poled lithium niobate (PPLN) crystal, which allows the generation of MIR temporal patterns through the DFG process. Such an indirect method of MIR pattern generation can avoid the bandwidth limitation of available MIR modulators \cite{Wu2024LSA}. To improve the DFG conversion efficiency, the average powers of the two fiber lasers are boosted to 1 W by using Er-doped and Yb-doped fiber amplifiers (EDFA and YDFA), which can produce a MIR beam with a maximum power of 1 mW. The MIR light is collimated by a calcium fluoride lens (LBTEK, BCX70613), before being spectrally filtered by a long-pass filter with a cutoff wavelength of 2.4 $\mu$m.

In our proof-of-principle demonstration, the temporal object is emulated by a MIR acousto-optic modulator (AOM) with an intensity transmission profile controlled by the AWG. The MIR modulated probe after the AOM is coupled into a single-mode fluoride fiber (Thorlabs, P33-32F-FC-1), which is then injected into an external-cavity-enhanced frequency upconverter based on the SFG process. In the upconversion unit, another PPLN crystal is placed at the waist of the cavity mode, which allows pump-enhanced spectral conversion of infrared radiation into the NIR band for subsequent sensitive detection \cite{Huang2021PR, Rodrigo2021LPR}. To enhance the SFG conversion efficiency, the CW single-longitudinal-mode pump is stabilized to the resonance of the optical cavity, which is realized by active adaptation of the position of the cavity mirror attached on a piezoelectric actuator (PZT). The feedback operation is performed with a digital locking unit based on the field-programmable gate array (FPGA) \cite{Liu2024APN}, which favors automatic monitoring and long-term stability. The upconverted signal at 810 nm is recorded by a silicon-based photodiode with high detection sensitivity. The implemented upconversion detector facilitates the room-temperature MIR single-photon detection \cite{Huang2021PR}, which is essential to implement the ultra-sensitive MIR computational TGI. The photodiode output is recorded by a real-time oscilloscope (LECROY, WaveSurfer 3034z). In the low-light scenario, a single-photon counting module (SPCM, Excelitas, SPCM-AQRH-54-FC) is employed to measure the photon-counting histogram with the help of a time-correlated single-photon counting (TCSPC) device (Qutools, quTAG). Finally, the temporal object is recovered from the correlation operation between the recorded integral intensity and the pre-programmed probe patterns. More details on the experimental setup are presented in Supplementary Note 1.

\section{Results and discussion}
\subsection{High-resolution MIR computational TGI}
\begin{figure*}[t!]
\includegraphics[width=0.75\textwidth]{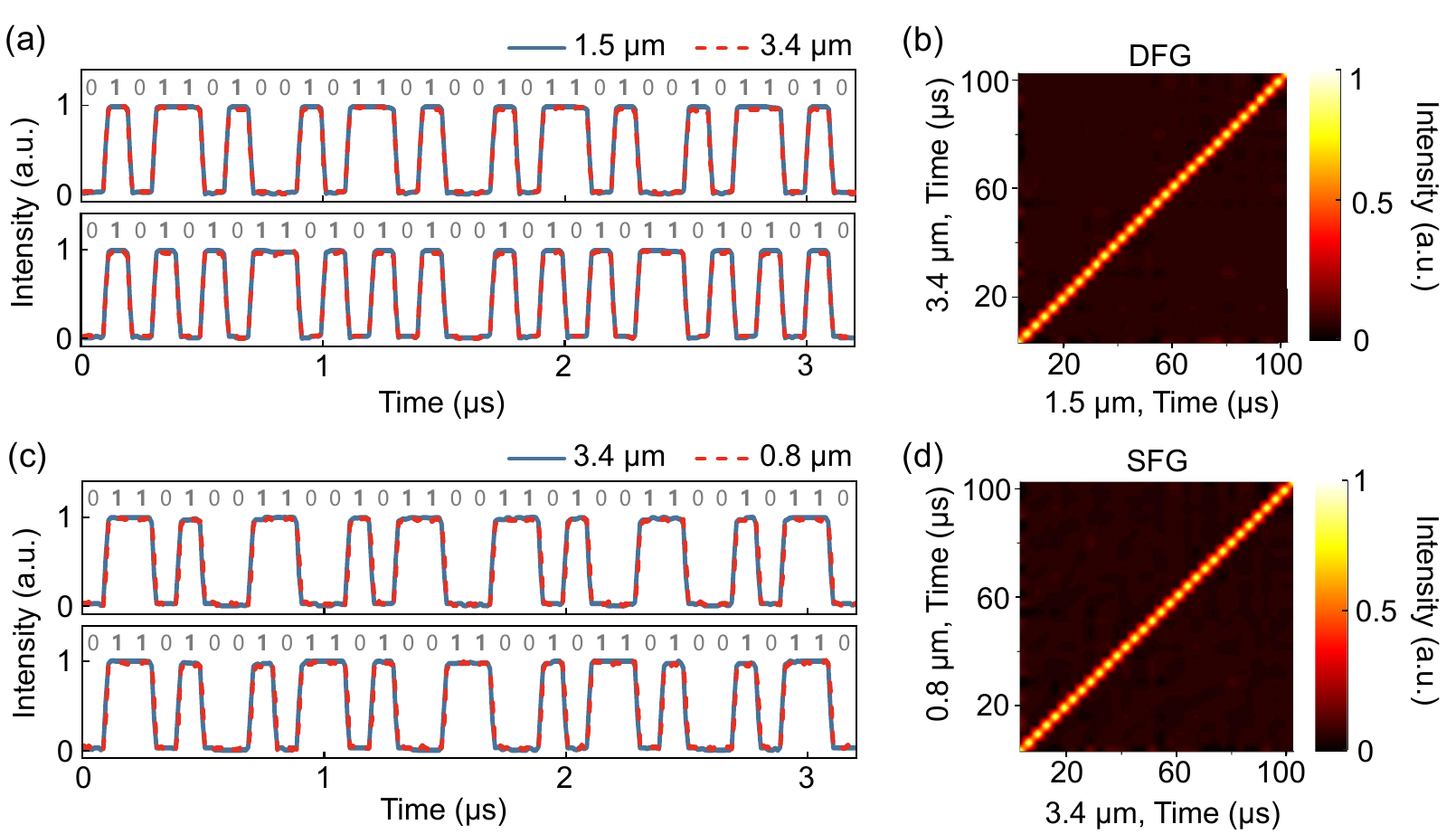}
\caption{High-fidelity transfer of the programmed temporal patterns during the DFG and SFG processes. (a) Comparison of time-resolved intensity profiles for the modulated light at 1550 nm and the down-conversion MIR patterns at 3.4 $\mu$m. Note that two representative bit sequences are taken from the 32-order Hadamard matrix. The binary encodings with nulls and ones are shown on the top of the temporal traces. The modulation rate of the EOM is set to be 10 Mbps, and the sequence duration is thus 3.2 $\mu$s. (b) Time-to-time intensity fluctuation cross-correlation between the dual-color temporal patterns, which are calculated over 32 sequences of distinct probing patterns. (c) Comparison of intensity profiles for the modulated light at 3.4 $\mu$m and the up-conversion MIR patterns at 810 nm. (d) Temporal fluctuation cross-correlation between the dual-color patterns.}
\label{fig2}
\end{figure*}

In our proposed scheme, two intermediate conversion steps are involved to facilitate both the preparation and detection of MIR temporal patterns. First, we start with verifying the high-fidelity information transfer between the NIR and MIR spectral regions. During the DFG process, the generated intensity for the MIR idler light is proportional to the product of the intensities of the two NIR signal and pump beams, as expressed by $I_\text{MIR}(t) \propto I_s(t) \times I_p(t)$. As a result, the MIR intensity profile directly follows the temporal patterns of the signal beam in the case of a CW pumping. The combination of NIR fast modulation and nonlinear spectral conversion offers an effective way to circumvent the deficiency of high-performance MIR modulators with high operation bandwidths and high extinction ratios \cite{Wu2019Optica, Wu2024LSA}.  Figure \ref{fig2}(a) presents the recorded waveforms for the signal and idler beams for two representative patterns in the 32-order Walsh Hadamard matrix. The modulation rate of the NIR pattern is set to be 10 Mbps, taking into consideration of the 5-MHz bandwidth of the MIR AOM as the temporal object. The NIR and MIR traces are detected by a high-speed InGaAs detector (Keyangphotonics  KY-PRM-3-500M-I-FC) with a bandwidth of 0.5 GHz and a HgCdTe detector (VIGO, UHSM-I-10.6) with a bandwidth of 0.7 GHz. The recorded dual-color waveforms are perfectly matched in the time domain. Moreover, the high-fidelity temporal information mapping is further manifested in the time-to-time intensity cross-correlation matrix, as given in Fig. \ref{fig2}(b). The correlation map is calculated by using a complete ensemble of 32 distinct bit sequences for the signal and idler beams. Similarly, the SFG process also implies a linear dependence on the MIR and SFG intensities via $I_\text{SFG}(t) \propto I_\text{MIR}(t) \times I_p(t)$. As shown in Figs. \ref{fig2}(c) and (d), we have experimentally confirmed the temporal correspondence between the MIR and upconverted beams, which lays the foundation to leverage the upconversion detection in the sensitive infrared TGI. In Supplementary Note 2, we have presented the measured temporal waveforms at signal, idler and SFG wavelengths for all the patterns in the Hadamard matrix, thus demonstrating the successful transfer of temporal encodings through the nonlinear wave-mixing operation.

\begin{figure*}[t!]
\includegraphics[width=0.7\textwidth]{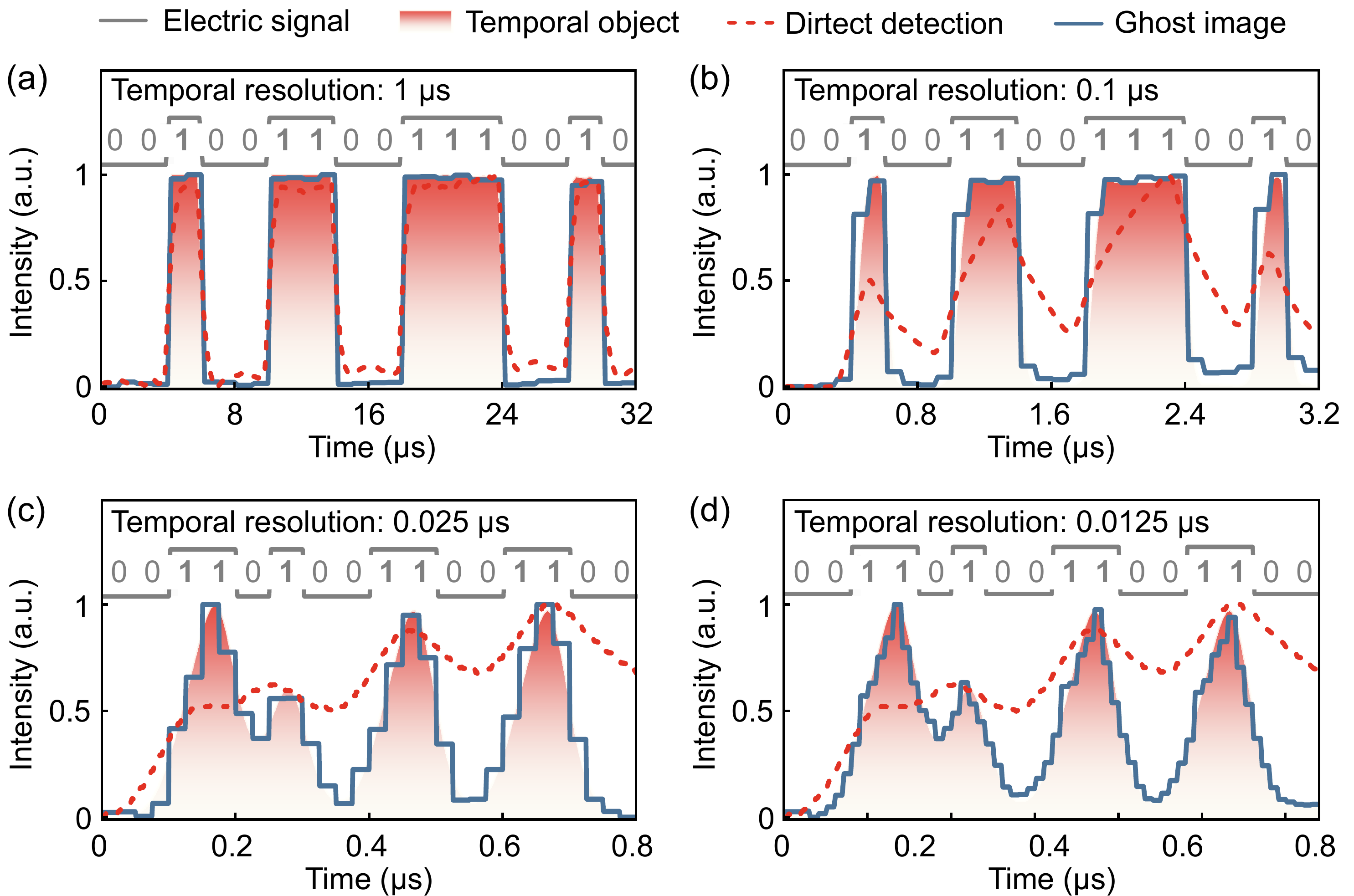}
\caption{MIR computational TGI for binary and gray temporal objects. (a) Ghost image in blue solid line is measured for a binary temporal object in the presence of structured probing patterns at a modulation rate of 1 Mbps. The gray solid line represents the electrical signal applied on the MIR AOM, which indicates the bit sequence as the encoding information. The ground truth of the temporal object in red shaded area is characterized by a MIR detector with a bandwidth of 0.7 GHz. As a comparison, the direct measurement in red dashed line is performed by a slow silicon detector with a bandwidth of 1 MHz. (b) Temporal imaging performances in the case of using probe patterns at a modulation rate of 10 Mbps. (c, d) Imaging performances for a gray temporal object with probe patterns at modulation rates of 40 Mbps (c) and 80 Mbps (d). Note that the structured patterns in (a-c) and (d) are from the 32- and 64-order Hadamard matrices, respectively.}
\label{fig3}
\end{figure*}

\begin{figure*}[t!]
\includegraphics[width=0.75\textwidth]{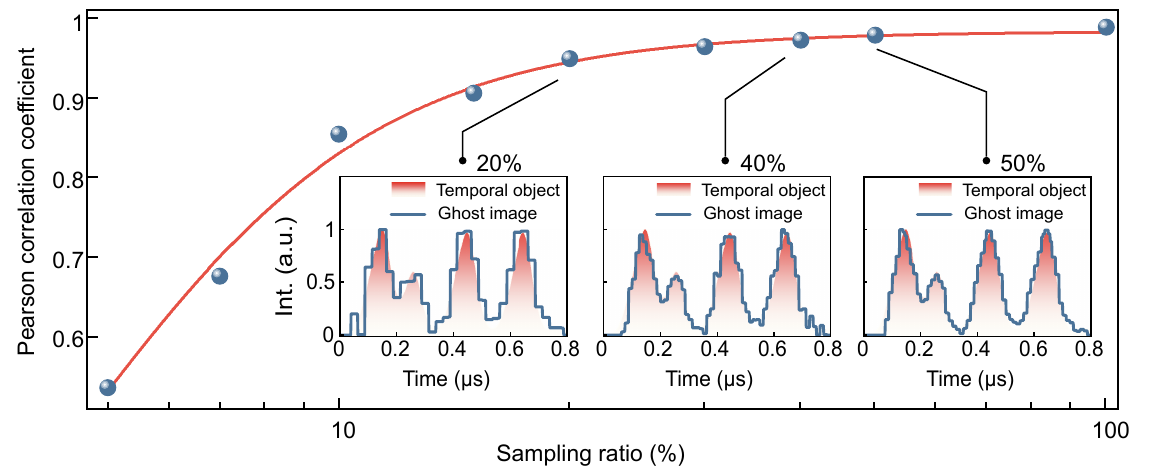}
\caption{Compressive MIR computation TGI. Reconstruction fidelity is evaluated by the Pearson correlation coefficient (PCC) as a function of
the sampling ratio. The solid line represents an empirical fit with the exponential function. Inset shows the reconstructed ghost images with sampling ratios of 20\%, 40\%, and 50\%. Note that the 128-order Hadamard probe pattern is used with a modulation rate of 160 Mbps, which corresponds to an interrogation window of 0.8 $\mu$s.}
\label{fig4}
\end{figure*}

Then, we turn to investigate the performance of the implemented MIR computational TGI system. In conventional TGI scheme based on random modulations, there inevitably exists information redundancy among the encoding patterns, thus necessitating a very large number of measurements for obtaining high-quality results \cite{Ryczkowski2016NP, Wu2019Optica}. In contrast, computational TGI uses a pre-programmed sequence of orthogonal temporal patterns, which allows not only substantial simplification of experimental setup, but also significant improvement of reconstruction efficiency \cite{Devaux2016Optica, Xu2018OE, Chen2021OL}. Specifically, we assume that the measured temporal object is given by $\mathcal{O} \in \mathbb{A}^{N \times1} $, where $N$ denotes the number of temporally resolved pixels. The measurement process can be described by $B = \Phi \mathcal{O}$, where $ B \in \mathbb{A}^{N \times 1}$ represents the integrated intensity measured by the bucket detector, and $\Phi \in \mathbb{A}^{N \times N}$ is the measurement matrix used in the computational TGI. In our experiment, $\Phi$ is a Hadamard matrix, which is a square matrix whose entries are either +1 or -1 and whose rows are mutually orthogonal.  Since the temporal intensity probes are non-negative, orthogonal modulation is achieved here through the differential signal acquisition method, which favors suppressing the intensity fluctuation of the light source \cite{Edgar2019NP, Zhao2021Optica}.  The generation process of the differential signal can be expressed as $\Phi_n = \Phi_{no} - \Phi_{ne}$, where $n$ represents the pattern sequence number, and $ \Phi_n$ ($ n = 1, 2, \ldots, N $) denotes the row sequences of the pre-programmed $N$-order Hadamard matrix. By replacing all -1 elements in $\Phi_n $ with 0 elements, $ \Phi_{no}$ is obtained. $\Phi_{ne}$ is the complementary matrix to $ \Phi_{no}$. One desirable feature of the Hadamard matrix is that the matrix inverse can be easily obtained through transposition, allowing simple and fast reconstruction of the temporal objects as expressed by $O = \Phi^{-1} B$. More discussion on the involved reconstruction algorithm is given in Supplementary Note 3.

Figure \ref{fig3} presents the TGI performance for interrogating binary and gray temporal objects. The binary target is prepared by modulating the MIR AOM with a square-wave waveform, which is directly measured by the HgCdTe detector to provide the ground truth of the targeted temporal profile. For a temporal object with a modulation rate of 0.5 Mbps, the probing patterns are configured to be operated at a two-fold frequency of 1 MHz. As shown in Fig. \ref{fig3}(a), the reconstructed waveform agrees well with the designated binary encodings. As a comparison, a direct detection is performed by simply using a silicon detector with a 3-dB bandwidth of 1 MHz in the absence of probe modulation. For a faster modulated temporal object, the direct measurement cannot precisely resolve the temporal structure as illustrated in Fig. \ref{fig3}(b). In this case, the TGI technique demonstrates the unique feature of achieving a high timing resolution with a slow detector. At a modulation rate of 10 MHz for the probe patterns, the temporal object can still be restored with a high quality. Intriguingly, the TGI supports the recovery of gray objects with both temporal and amplitude features. To this end, we increase the modulation rate of the electrical waveform to a value beyond the bandwidth of the MIR AOM, which results in a series of filtered peaks with varying slopes and heights. Figure \ref{fig3}(c) illustrates the temporal object in the case of using electrical codings at a rate of 20 Mbps. At a modulation rate of 40 Mbps for the probe patterns, the object profile can be well identified in terms of temporal and amplitude structures. Moreover, more accurate measurement with a higher temporal resolution is conducted by increasing the probing rate to 80 Mbps, which leads to doubling the number of sampling points to obtain a high-resolution waveform as shown in Fig. \ref{fig3}(d).

\begin{figure*}[t!]
	\includegraphics[width=0.9\textwidth]{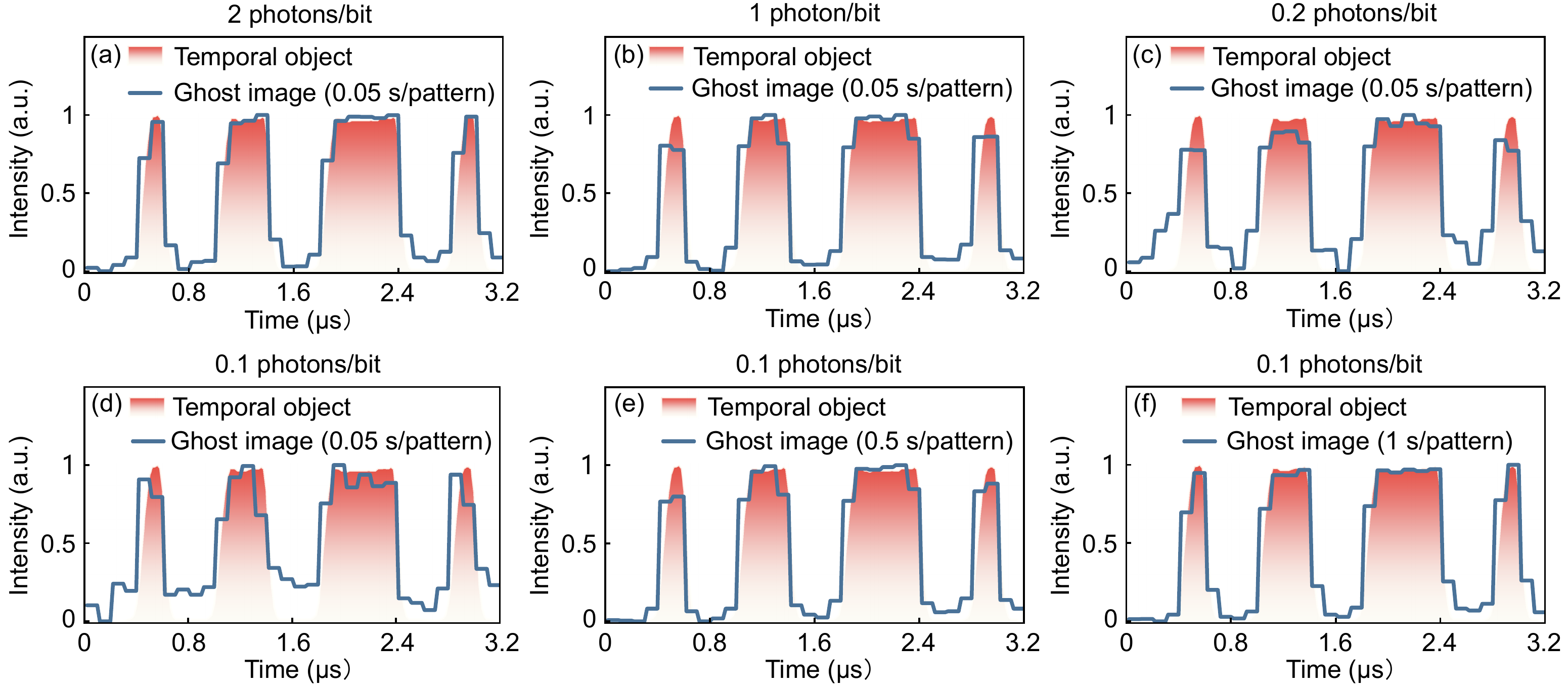}
	\caption{Single-photon MIR computational TGI. (a-c) Performance comparison for the reconstructed MIR ghost images under illumination fluxes of 2 photons/bit (a), 1 photon/bit (b), and 0.2 photons/bit (c). The acquisition time is set to be 0.05 s for each pattern. (d-f) Reconstructed MIR ghost images in the case of an integration time of 0.05 s (d), 0.5 s (e), and 1 s (f) for each pattern, respectively. The light flux for the code sequence is fixed to be 0.1 photons/bit.}
	\label{fig5}
\end{figure*}

\subsection{Compressive MIR computational TGI}
In the following, we proceed to demonstrating compressive MIR ghost imaging in the time domain, which provides an efficient way to recover high-resolution temporal profiles at a sampling rate below the Nyquist-Shannon limit. In the above results, the number of measurements is required to be identical to the pixel number of the reconstructed temporal trace. In contrast, a sophisticated strategy based on the compressive sensing algorithm allows the reconstruction of signals from fewer measurements without significant loss of information, which is made possible due to the inherent sparsity of most natural objects \cite{Edgar2019NP, Zhao2021Optica}. In the compressive sensing approach, $M$ ($M < N$) measurements are used to perform the signal recovery, as expressed by $B = \Phi \mathcal{O} = \Phi \Psi s = \Theta s$, where $B \in \mathbb{A}^{M \times 1}$,  $\Phi \in \mathbb{A}^{M \times N}$, and $ \Theta = \Phi \Psi \in \mathbb{A}^{M \times N} $. Note that $s \in \mathbb{A}^{N \times 1} $ is the sparse vector that contains the projection of the signal in the sparse sampling basis $ \Psi \in \mathbb{A}^{N \times N}$. With the total variation minimization algorithm, the signal can be stably recovered with the compressive measurements \cite{Zhao2021Optica, Sun2024LPR}.

In the experiment, $M$ rows from the Walsh-ordered Hadamard matrix are randomly selected to constitute the measurement matrix \cite{Sun2024LPR}. By correlating the projected patterns and corresponding intensities recorded on the detector, an $N$-element temporal waveform can be reconstructed via the total variation augmented Lagrangian (TVAL3) algorithm. Further details of the reconstruction procedures are provided in Supplementary Note 3. The sampling ratio is generally used to quantify the data compression capability during the signal acquisition, which is defined by the ratio between the number of measurements and the number of resolvable elements in the reconstructed waveform. Meanwhile, the reconstruction quality can be evaluated by the Pearson correlation coefficient (PCC), which indicates the similarity between the recovered signal and the ground truth for the temporal object. Figure \ref{fig4} presents the calculated PCCs as a function of the sampling ratio. High-fidelity reconstruction with a PCC above 85\% is permitted with a small number of realizations down to 10\%. The PCC can be improved to be over 95\% by resorting to a larger sampling ratio above 20\%. The implemented compressive sensing operation significantly reduces the acquisition time, which favors a faster signal analysis in comparison to the time-domain raster scanning scheme \cite{Edgar2019NP}.

\begin{figure}[b!]
	\includegraphics[width=0.8\columnwidth]{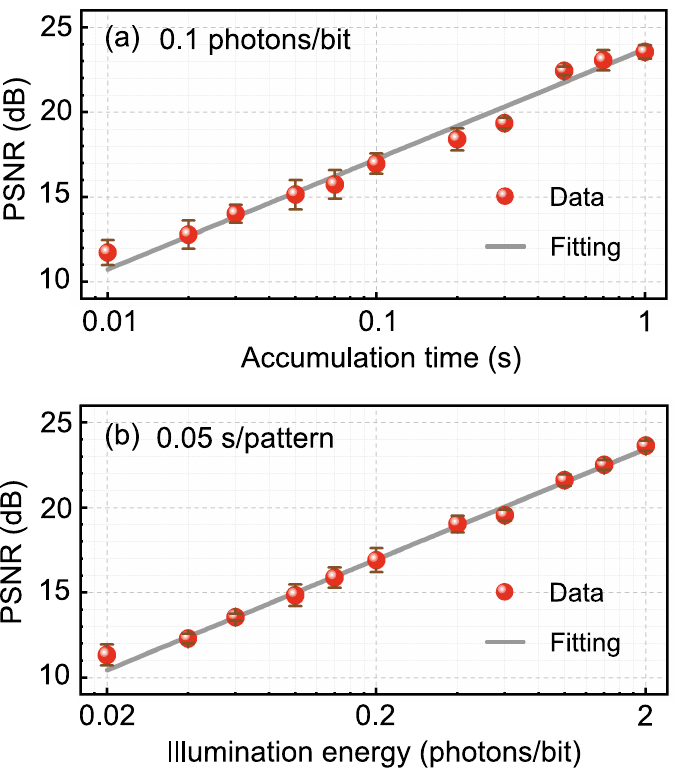}
	\caption{Detection sensitivity for the MIR computational TGI. (a) Peak signal-to-noise ratio (PSNR) varies with the increase of accumulation time for each pattern under a constant illumination flux of 0.1 photons/bit. (b) PSNR as a function of the incident bit energy onto the temporal object at a fixed accumulation time of 0.05 s for each pattern. The error bars denote  standard deviations for five repetitive experiment measurements. The modulation rate of the MIR probe is set to be 10 Mbps.}
	\label{fig6}
\end{figure}

 \begin{figure*}[t!]
	\includegraphics[width=0.75\textwidth]{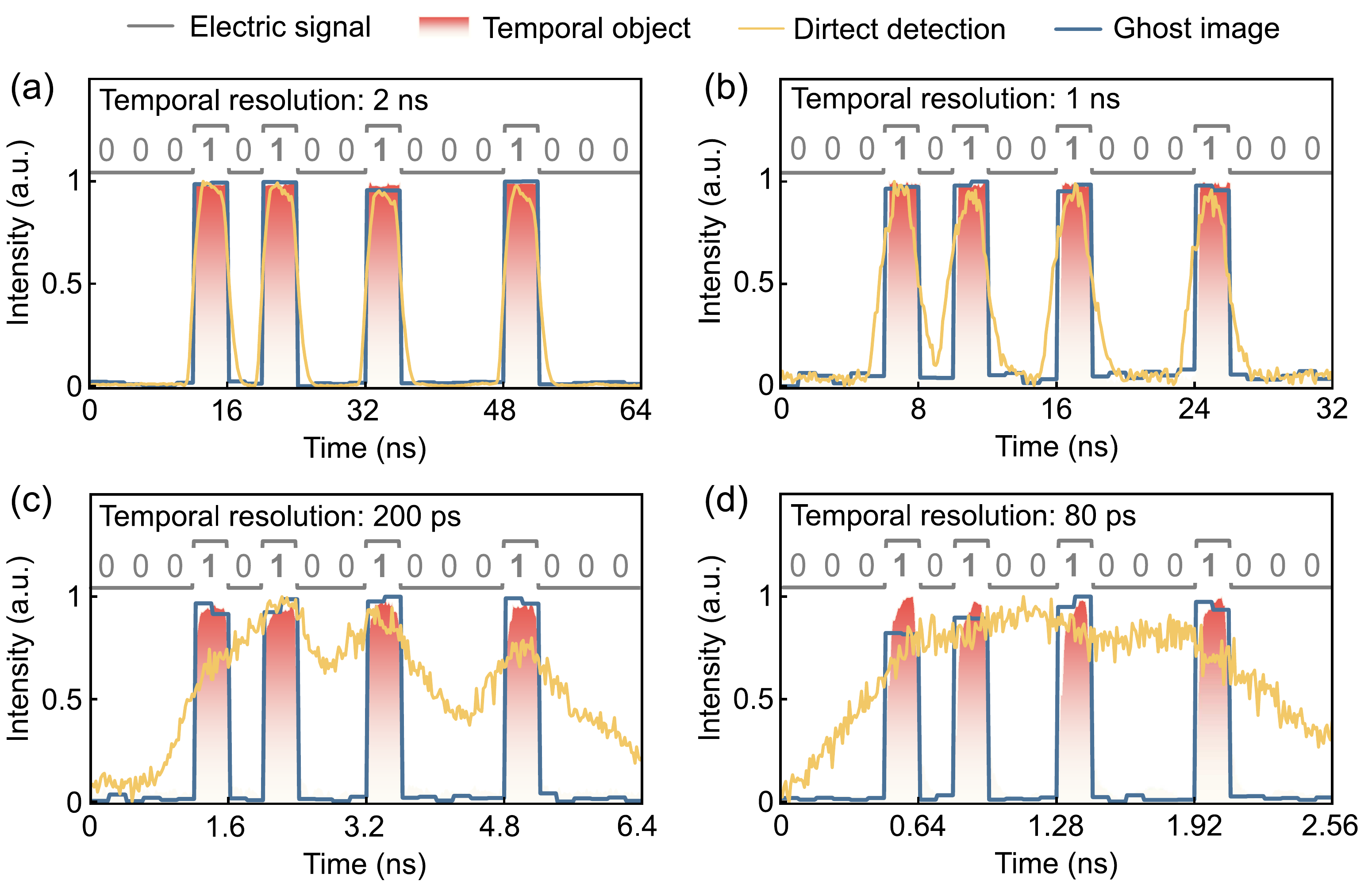}
	\caption{High-precision MIR single-photon computational TGI. (a) Measured ghost image in blue solid line for a binary temporal object in the case of using structured probing patterns at a modulation rate of 500 Mbps. The gray solid line denotes the designated bit sequence for the encoding information. The red shaded area represents the ground truth of the temporal object. As a direct comparison, the yellow solid line shows the recorded photon-counting histogram by the conventional time-correlated single-photon counting (TCSPC) technique. (b-d) Temporal imaging performances for for a binary object with probe patterns at modulation rates of 1 Gbps (b), 5 Gbps (c) and 12.5 Gbps (d), respectively. Note that the MIR temporal object is emulated by the down-conversion replica of the fast-modulated light profile from the near-infrared EOM. The prepared structured patterns are prepared with the 32-order Hadamard matrix. The achieved temporal resolution of 80 ps surpasses the timing-jitter limitation of 846 ps for the used single-photon detector in the experiment.}
	\label{fig7}
\end{figure*}

\subsection{Single-photon MIR computational TGI}
Finally, we focus on the investigation of the MIR computational TGI at the single-photon level. The single-photon temporal imaging is manifested by an extremely weak flux for the incident MIR signal intensity. To this end, the MIR power is adjusted by selecting an appropriate combination of neutral density attenuators and precisely changing the pump power in the DFG unit, which facilitates both the coarse and fine tuning of the attenuation \cite{Huang2021PR}. The accurate power calibration within a large dynamic range is crucial to characterize the subsequent low-light-level measurement. Accordingly, the detector is replaced by a SPCM based on a Geiger-mode silicon avalanche photodiode. The SPCM has a detection efficiency of 60\% at 810 nm, and is specified with a timing jitter of 846 ps. Different from an analog detector, the single-photon detector produces a series of pulses corresponding to incident photons, thus requiring a frequency counter to record the event count within a customized integration time. To emulate the photon-starved scenario, the MIR waveform before passing the temporal object is attenuated to 2, 1 and 0.2 photons/bit as shown in Figs. \ref{fig5}(a-c), respectively. The time slot for each bit is 100 ns, which results in a sequence time of 3.2 $\mu$s for 32 binary codes. The photon counts are measured within the window of the bit sequence, and averaged over an accumulation time of 0.05 s/pattern. The reconstructed ghost images agree well with the ground truth of the temporal object. In the presence of a weaker light flux at 0.1 photons/bit, the signal-to-noise ratio (SNR) can be improved by increasing the integration time, as shown in Figs. \ref{fig5}(d-f). Particularly, the reconstructed quality in Fig. \ref{fig5}(f) is close to the one obtained in Fig. \ref{fig5}(a) due to identical photons collected for each pattern.

To quantitatively evaluate the accuracy of the retrieved ghost images, the peak signal-to-noise ratio (PSNR) is calculated between the directly measured and reconstructed temporal sequences (see Supplementary Note 4 for the analytical definition) \cite{Xu2018OE, Wu2024LSA}. Figure \ref{fig6}(a) presents the measured PSNR as a function of the accumulation time for each projected pattern. The incident photon flux is set to be 0.1 photons/bit for the MIR probe sequence. In the case of a constant integration time of 0.05 s/pattern, the PSNR increases with a higher illumination power, as shown in Fig. \ref{fig6}(b). Furthermore, we have compared the performances with measurement matrices based on the raster scanning and Hadamard masks, as given in Supplementary Note 4. In the point-scanning scheme, the detected signal is from a single time slot for each bit, which results in a noisy waveform especially under the photon-starved illumination. In contrast, the temporal encoding modality favors the multiplex advantage to increase the light collection efficiency in each measurement \cite{Sun2024LPR}. The resultant SNR improvement relies on the similar mechanism in the single-pixel imaging \cite{Edgar2019NP}.

Typically, the single-photon waveform is temporally analyzed with the TCSPC device, where the time intervals between events and the trigger are registered to produce photon-counting histograms for the arrival times of individual photons. Although the time-stamping precision of the photon counter can reach the picosecond level, the temporal resolving power for the TCSPC system is usually limited to sub-nanosecond due to the timing jitter of the single-photon detector. To demonstrate the full potential of the infrared TGI system, the temporal object here is emulated by modulating the NIR EOM. In this case, the targeted temporal profile is superimposed with the structured encodings via the ``and"-gate operation. Consequently, high-speed MIR bit sequences can be prepared with a temporal resolution of 80 ps, as technically governed by the pulse pattern generator in the experiment. Figure \ref{fig7} presents the performance comparison between the direct measurement and computational TGI for characterizing sequence durations from 64 to 2.56 ns. As the time slot of each bit becomes shorter than 200 ps, the directly recorded histogram by the TCSPC cannot properly reveal the actual shape of the binary encodings as shown in Figs. \ref{fig7}(c) and (d). In contrast, the computation TGI allows one to successfully recover the bit sequence with a temporal resolution of 80 ps, which is well beyond the timing jitter of 846 ps for the single-photon detector. In further combination of more advanced pulse generators and light modulators with bandwidths over 100 GHz, the presented TGI paradigm paves the way toward examining time-dependent processes with picosecond temporal resolutions, which would simulate wide-ranging applications like ultrafast pump-probe measurements \cite{Julsgaard2020PR} and time-domain spectroscopy \cite{Zhao2021Optica}.

\section{Conclusion}
Over the past decade, the TGI technique has emerged to provide an appealing solution to mitigate the stringent bandwidth requirement for high-speed signal detection \cite{Ryczkowski2016NP, Devaux2016Optica}. However, the implementation of single-photon TGI necessitates sensitive detectors and ultrafast modulators for measuring and pre-programming the probing intensity patterns, which are not available in all spectral regions, particularly beyond the visible or NIR wavelengths. Recently, the pioneering demonstration of MIR TGI has been reported albeit with the detection sensitivity far away from the single-photon level \cite{Wu2024LSA}. So far, MIR single-photon TGI has not yet been demonstrated, which urgently calls for effective solutions to realize high-speed intensity modulation and high-sensitivity photon counting in the MIR region.

In this work, we have implemented for the first time a MIR computational TGI system at the single-photon level, which addresses the challenges of high-speed pattern generation and high-sensitivity photon detection in the MIR region. In contrast to previous infrared TGI schemes \cite{Wu2019Optica, Wu2024LSA}, the presented approach eliminates the stringent requirements on the direct MIR modulators and detectors through the high-fidelity nonlinear spectral conversion. In this architecture, high-performance NIR modulators and detectors can be leveraged to realize MIR waveform analyses with desirable features of single-photon sensitivity, picosecond time resolution, and room-temperature operation. As summarized in Supplementary Note 5, the achieved detection sensitivity here represents a significant landmark among previously reported TGI instantiations. Moreover, the presented single-photon TGI scheme makes it possible to obtain superior temporal resolutions beyond the intrinsic timing jitter of the photon counter, which opens up novel possibilities in fast and sensitive MIR measurement.

Notably, the involved frequency converters for MIR generation and detection can be simplified with the single-pass variants based on PPLN waveguides, thus favoring more compact footprint and easier system implementation. Moreover, there exist several improvements to go beyond the achieved performances. First, resorting to non-oxide nonlinear crystals like silver gallium sulfide (AgGaS$_2$) and gallium phosphide (GaP), the proposed generic approach can be readily implemented in long-infrared \cite{Rodrigo2021LPR} or terahertz \cite{Zhao2021Optica} spectral regions, where high-performance detectors and optical modulators are currently difficult to access. Second, benefitting from the state-of-the-art telecom technology, the operation bandwidth up to 100 GHz is feasible to approach a temporal resolution of 10 ps \cite{Shi2024NP, Wey1995JLT}. We note that programmable generation of NIR pulsed sequences with a sub-picosecond precision has recently been demonstrated by using a digital micromirror device and a femtosecond light source \cite{Zhao2021Optica}, which could further enhance the temporal resolving ability of the MIR computational TGI system. Third, in Supplementary Note 6, we propose a feasible scheme to realize the MIR single-shot computational TGI by using the dense wavelength-division multiplexing (DWDM) technology. The single-shot operation is favorable for capturing transient dynamic processes in real time without the need to repeat the event \cite{Devaux2017JO, Devaux2016Optica, Tang2018PTL}, which not only allows the detection of asynchronous and non-reproducible signals, but also paves a novel path toward high-capacity free-space MIR optical communications \cite{Zou2022NC}.
\newline
\newline

\section*{Acknowledgements}
This work was supported by the Shanghai Pilot Program for Basic Research (TQ20220104); the National Natural Science Foundation of China (62175064, 62235019, 62035005); the Innovation Program for Quantum Science and Technology (2023ZD0301000); the Natural Science Foundation of Chongqing (CSTB2023NSCQ-JQX0011, CSTB2022TIAD-DEX0036); the Shanghai Municipal Science and Technology Major Project (2019SHZDZX01); and the Fundamental Research Funds for the Central Universities.

\section*{Conflict of Interest}
The authors declare no conflict of interest.

\section*{Supporting Information}
Supporting Information is accompanied to present more details on the experimental settings, data processing, and possible extensions. 

\section*{Data Availability Statement}
The data that support the findings of this study are available from the corresponding author upon reasonable request.

\section*{Keywords}
temporal ghost imaging; mid-infrared detection; single-photon detection; compressive sensing measurement; frequency upconversion detection


\end{document}